\documentclass[12pt,preprint]{aastex}

\newcommand{\dmm}{\mbox{$\Delta$m$_{15}(B)$}}

\shorttitle{U-band photometry}
\shortauthors{Krisciunas et al.}

\begin{document}

\title{Fixing the $U$-band photometry of Type Ia supernovae\altaffilmark{1}}

\author{Kevin Krisciunas,\altaffilmark{2,3}
Deepak Bastola,\altaffilmark{2}
Juan Espinoza,\altaffilmark{4}
David Gonzalez,\altaffilmark{4}
Luis Gonzalez,\altaffilmark{5}
Sergio Gonzalez,\altaffilmark{5}
Mario Hamuy,\altaffilmark{6}
Eric Y. Hsiao,\altaffilmark{5}
Nidia Morrell,\altaffilmark{5}
Mark M. Phillips,\altaffilmark{5}
and Nicholas B. Suntzeff\altaffilmark{2,3}
}
\altaffiltext{1}{Based in part on observations taken at the Cerro Tololo 
Inter-American Observatory, National Optical Astronomy Observatory, which is 
operated by the Association of Universities for Research in Astronomy, Inc. (AURA) 
under cooperative agreement with the National Science Foundation.}

\altaffiltext{2}{Texas A\&M University, Department of Physics and Astronomy, 4242 TAMU,
College Station, TX 77843-4242; {krisciunas@physics.tamu.edu} 
{suntzeff@physics.tamu.edu} }

\altaffiltext{3}{George P. and Cynthia Woods Mitchell Institute for
Fundamental Physics \& Astronomy, 
Texas A\&M University, Department of Physics, 4242 TAMU,
College Station, TX 77843-4242}

\altaffiltext{4}{Cerro Tololo Inter-American Observatory, Casilla
  603, La Serena, Chile; {jespinoza@ctio.noao.edu}}

\altaffiltext{5}{Las Campanas Observatory, Casilla 601, La Serena, Chile;
  {hsiao@lco.cl} {nmorrell@lco.cl} {mmp@lco.cl} }

\altaffiltext{6}{Universidad de Chile, Departamento de Astronom\'{i}a, Casilla 36-D,
Santiago, Chile; {mhamuy@das.uchile.cl}}

\begin{abstract}
  
We present previously unpublished photometry of supernovae 2003gs and 
2003hv.  Using spectroscopically-derived corrections to the $U$-band 
photometry, we reconcile $U$-band light curves made from imagery with the 
Cerro Tololo 0.9-m, 1.3-m and Las Campanas 1-m telescopes.  Previously, 
such light curves showed a 0.4 mag spread at one month after maximum 
light.  This gives us hope that a set of corrected ultraviolet light 
curves of nearby objects can contribute to the full utilization 
of rest frame $U$-band data of supernovae at redshift $\sim$0.3 to 0.8.  
As pointed out recently by Kessler et al. in the context of the Sloan 
Digital Sky Survey supernova search, if we take the published $U$-band 
photometry of nearby Type Ia supernovae at face value, there is a 0.12 
mag $U$-band anomaly in the distance moduli of higher redshift objects.  
This anomaly led the Sloan survey to eliminate from their analyses 
all photometry obtained in the rest frame $U$-band.  The Supernova
Legacy Survey eliminated observer frame $U$-band photometry, which is
to say nearby objects observed in the $U$-band, but they used 
photometry of high redshift objects no matter in which band the photons
were emitted.

\end{abstract} 


\section{Introduction}  

Type Ia supernovae (SNe) are very useful standardizable candles for
extragalactic astronomy and observational cosmology.  A Type Ia SN
is often thought to be a carbon-oxygen white dwarf that approaches the
Chandrasekhar limit of 1.4 M$_{\sun}$ owing to mass transfer from
a close main sequence stellar companion or giant star \citep{Whe_Ibe73, Liv00}.
Some Type Ia SNe might be mergers of two white dwarfs 
\citep{Web84, How11, Sch_Pag12}.  These supernovae provided
the first observational evidence that the expansion of the universe
is accelerating \citep{Rie_etal98, Per_etal99}; three members of
the two key groups that carried out this work were awarded the
Nobel Prize in Physics in 2011.

Three of the most important datasets containing $U$-band photometry of 
Type Ia SNe are the ``CfA2 sample'' of $UBVRI$ photometry of 44 objects 
by \citet{Jha_etal06}, the ``CfA3 sample'' of 185 objects \citep{Hic_etal09},
and the recent ``CfA4 sample'' of 94 objects \citep{Hic_etal12}.  
The ``CfA2 sample'' was the first large dataset containing 
$U$-band light curves of Type Ia SNe.  The ``CfA4'' sample contains
$U$-band data of 14 objects and $u^{\prime}$ photometry 12 other
objects.  These datasets were produced by astronomers at the Harvard-Smithsonian 
Center for Astrophysics.

Of the \citet{Hic_etal09} sample, 31 objects have $U$-band maxima,
values of the decline rate parameter \dmm, and redshifts greater than
$z$ = 0.01.  The $U$-band Hubble diagram shows a scatter of about
$\pm$ 0.25 mag. Some of the scatter may be due to the asymmetric nature of
some of these explosions, in which case what we
see is a function of the viewing angle \citep{Mae_etal10}.  Some
of the scatter may be due to incorrect extinction corrections for dust along
the line of sight, or differences amongst the various $U$-band filters used.
The larger scatter in the $U$-band Hubble diagram could also be due in part
to spectroscopic differences that correlate with metallicity, galaxy type,
and redshift.  \citet{Fol_etal08a}, \citet{Fol_etal12}, and 
\citet{Mag_etal12} have provided evidence that
higher redshift Type Ia SNe have different spectral energy distributions in
the ultraviolet than nearby objects.

By comparison, the scatter in the Hubble diagram
is $\pm$ 0.15 mag or better for other optical
or near-IR bands \citep{Phi_etal99,Fol_etal10,Kat_etal12,Kri12}.
If we observe Type Ia SNe beyond a redshift of $z \approx$ 0.03,
the effect of peculiar velocities diminishes, and the observed
scatter of the absolute magnitudes is reduced to $\pm$ 0.12
mag in the near-IR $J$- and $H$-bands \citep{BN_etal12}. This
corresponds to a $\pm$ 6 percent uncertainty in distance.

For an object at, say, redshift 0.7, the photons we observe in the 
$R$-band ($\lambda \approx 0.65 \mu$m)were emitted at ultraviolet wavelengths.  
Medium deep SN surveys 
such as ESSENCE \citep{Woo_etal07}, the Supernova Legacy Survey 
\citep[SNLS,][]{Con_etal11}, and the Sloan Digital Sky Survey 
\citep[SDSS,][]{Kes_etal09} include a significant percentage of restframe 
ultraviolet observations.  A tough problem arises.  \citet{Kes_etal09} 
show that if we include nearby objects observed in the $U$-band
along with the higher redshift objects whose $U$-band light has
been redshifted to longer wavelength passbands,
there is a 0.12 mag shift in the distance moduli of the 
high-redshift sample, which leads to a 0.3 shift of the cosmic equation 
of state parameter $w$.  This is a huge shift!  SDSS-II 
decided to eliminate from analysis all photometry that 
originated in the rest-frame UV.  SNLS, on the other hand, used photometry
of high redshift objects even if the photons were emitted in the $U$-band,
but eliminated $U$-band photometry of nearby objects.  We note the
0.12 mag anomaly in the distance moduli may be due to a $\sim$0.05 mag
shift in the $U$-band photometry of objects in the CfA2 sample
\citep[][p. 67]{Kes_etal09}; such
a systematic error in the CfA2 $U$-band magnitudes is actually smaller
than the typical rms scatter of fully corrected $U$-band light curves
(see below).  (Note that the CfA3 and CfA4 samples were published after
the SDSS analysis.)

In \S10.1.3 of their paper \citet{Kes_etal09} list five possibilities
to explain the $U$-band anomaly: 1) redshift dependent flux; 2) selection
effects for the nearby sample; 3) problems with the lightcurve fitting
model(s); 4) photometric calibration errors; and 5) differences in the
UV spectral energy distributions of the supernovae.  We remind
the reader that most photometric calibration errors are of two kinds:
1) problems with the standard stars; or 2) insufficient knowledge of
the effective passbands, leading effectively to multiple ``systems''.   
There appears to be unexplained variables in the CfA2 sample.
The present paper primarily addresses the passband issue.  

If we combine observations of a particular SN obtained with different 
cameras on different telescopes (or even the same camera on the same 
telescope, but with physically different filters), there can be significant 
systematic differences in the light curves, particularly in the $U$-band. 
Figure \ref{u_uncorrected} shows that one month after maximum light the 
$U$-band data has spread out by 0.4 mag for two particular objects.  This 
is the reason \citet{Kri_etal09} did not publish the $U$-band 
photometry of SN 2003gs obtained with the Las Campanas Observatory (LCO) 
1-m telescope. At that time we could not resolve the telescope to 
telescope differences.  The one hopeful feature of the uncorrected 
$U$-band light curves of SNe 2003gs and 2003hv is that the CTIO 0.9-m and 
LCO 1-m photometry is offset from the CTIO 1.3-m photometry by about the 
same amounts for each object.

In this paper we show that spectroscopically-derived corrections
to $U$-band photometry can effectively cure the problem that is
so obvious in Figure \ref{u_uncorrected}.
If appropriate S-corrections are applied to a specific set of
$\sim$30 $U$-band light curves of nearby Type Ia SNe, we might
be able to resolve the $U$-band anomaly that so affected the
analysis of SDSS-II.  

\section{Photometric reduction and the method of S-corrections}

On some given night the
standardized magnitudes of \citet{Lan92} standards may be related to
instrumental magnitudes and instrumental colors as follows:

\begin{equation}
U \; = \; u \; - k_U X \; + ct_U (u-b) + zp_U \; ,
\end{equation}

\begin{equation}
B \; = \; b \; - k_B X \; + ct_B (b-v) + zp_b \; ,
\end{equation}

\begin{equation}
V \; = \; v \; - k_V X \; + ct_V (b-v) + zp_V \; ,
\end{equation}

\begin{equation}
R \; = \; r \; - k_R X \; + ct_R (v-r) + zp_R \; ,
\end{equation}

\begin{equation}
I \; = \; i \; - k_i X \; + ct_I (v-i) + zp_I \; ,
\end{equation}

\parindent = 0 mm

where $k$'s are atmospheric extinction coefficients, 
measured in magnitudes per airmass, X is the airmass
(basically the secant of the zenith angle), ct's are color
terms, and zp's are zero points.  Typical extinction  
coefficients at Cerro Tololo or Las Campanas are
$k_U$ = 0.51, $k_B$ = 0.26, $k_V$ = 0.15, $k_r$ = 0.11, and $k_i$ = 0.06
magnitudes per airmass. The zero points and extinction
vary from night to night, even if the nights are photometric, and
the $U$-band parameters can even vary over the course of a single night.

\parindent = 9 mm

In Equations 1 through 5 the particular instrumental color used should
include the band which is being transformed to catalog magnitudes.
For example, we use $v-r$ in Equation 4, but could have just as easily
used $r-i$.  It would not make sense to use either of these instrumental
colors for Equation 1, however.  

Mean color terms for the CTIO 1.3-m telescope and ANDICAM from August
2003 through October 2003 were 
ct$_U$ = $-$0.109,
ct$_B$ =   +0.054,
ct$_V$ = $-$0.040,
ct$_R$ =   +0.004,
ct$_I$ = $-$0.067, with uncertainties of $\pm$ 0.004.
For the CTIO 0.9-m telescope ct$_U$ = +0.119 $\pm$ 0.007, and
for the LCO 1-m telescope ct$_U$ = +0.185 $\pm$ 0.011 during this time.

The method of spectroscopically derived corrections to the photometry
of SNe was first laid out by \citet{Str_etal02} and \citet{Kri_etal03}.
Basically, we take the nominal filter profiles, determined in the lab
or provided by a manufacturer, and multiply those by a number of other
functions of wavelength to account for transmission of the light
through the Earth's atmosphere, reflection off of aluminum coated
mirrors, the effect of any field lenses or a dichroic beamsplitter,
and the quantum efficiency of the chip.  The nominal effective
filter profile is then shifted arbitrarily in wavelength so that
synthetic magnitudes of standard stars reproduce the photometric
color terms that one measures directly doing photometry with
a particular telescope and camera. \citet{Kri_etal03} and
\citet{Kri_etal04} show the 
$BVRI$ S-corrections for the CTIO 1.3-m and 0.9-m telescopes.
The numerical values of these corrections are added to appropriate
standardized magnitudes of Equations 1 to 5 to correct 
the photometry to what we would have obtained, had we observed with the
\citet{Bes90} filters.

Figure \ref{u_trans} shows the effective filter profiles of
the $U$-band filters of the CTIO 0.9-m, CTIO 1.3-m, LCO 1-m telescopes
and the \citet{Bes90} filter prescription (with appropriate atmospheric
and CCD chip quantum efficiciencies accounted for).  Note how blue
the $U$-filter of the CTIO 1.3-m telescope camera (ANDICAM) is compared
to those of the CTIO 0.9-m and LCO 1-m telescopes.  

To calculate the $U$-band S-corrections we used spectra of 50 standard stars by
\citet{Str_etal05} and calculated synthetic magnitudes using
a script written by one of us (N. B.S.) which
runs in the {\sc iraf} environment.\footnote[7]{{\sc iraf} is distributed by 
the National Optical Astronomy Observatory, which is operated by the Association 
of Universities for Research in Astronomy, Inc., under cooperative
agreement with the National Science Foundation (NSF).} At the short
wavelength end, if spectra do not extend to the short wavelength limit of 
a filter, the flux points are set to zero.  This also holds for spectra
that do not extend to the long wavelength limit of a filter.
These standard star spectra were extended to $\lambda$ = 3100 \AA\ 
in the blue by means of Kurucz stellar atmosphere code, as modified
by W. Vacca and P. Massey.  See \S3 of \citet{Str_etal05}.

$U$-band S-corrections for four ``normal'' Type Ia SNe are shown in Figure \ref{u_scorr}. 
For subsequent purposes we adopt the low order polynomial
fits to the individual points to correct $U$-band photometry.  Typical uncertainties
of the $U$-band S-corrections are $\pm$ 0.03 to 0.04 mag.

We should consider what systematic errors there may be in our synthetic magnitudes 
and S-corrections owing to SN spectra not extending to the short wavelength limits 
of the effective filter profiles.  Using the \citet{Hsi_etal07} SN template 
spectra truncated at 3200 \AA\ (and not truncated) we can perform some 
experiments.  Owing to truncation at the short wavlength end of the spectra, the 
systematic errors in the S-corrections are greatest for the ANDICAM $U$-band 
filter, as it is the bluest shown in Figure \ref{u_trans}. Some of our $U$-band 
S-corrections at T($B_{max}$) may be too negative by 0.06 mag, diminishing to 0.03 
to 0.04 mag afterwards.  For the CTIO 0.9-m and LCO 1-m, our S-corrections may be 
too positive by 0.01 to 0.02 mag.  These systematic errors are noticeably smaller 
than the typical scatter of single-telescope $U$-band photometry of Type Ia SNe 
(typically $\pm$ 0.07 mag or greater).

In Figure \ref{u_scorr} we also show some S-corrections calculated with the Hsiao 
templates.  For this purpose the templates were warped to match the natural system 
magnitudes of SNe~1999ee (from the CTIO 0.9-m telescope) and 2002bo (from the 
Yale-AURA-Lisbon-Ohio telescope at CTIO).  In the case of SN~1999ee the template
spectra still have excess flux in the two humps at 3050 and 3450 \AA\ relative
to the double hump at 3900 to 4050 \AA, leading to S-corrections that do not
match the values based on actual spectra of SN~1999ee prior to $t$ = 15 d.
The agreement for SN~2002bo from $0 \leq t \leq 15$ d is good, however.
We conclude that warped \citet{Hsi_etal07} templates can be used in the
$U$-band, but with caution.  

For SN 1999ee we used spectra from \citet{Ham_etal02}; the earliest spectrum
extends to $\lambda$ = 2965 \AA\ while the others extend to 3260 \AA.  
For SN 2001ay we used the ``CfA set'' of spectra reduced by T. Matheson; these
are spectra obtained with the Fred L. Whipple Observatory 1.5-m telescope and the 
Multiple Mirror Telescope \citep {Kri_etal11}. For
SN 2001el we used spectra obtained by Peter Nugent with the {\em Hubble Space
Telescope} and previously used by \citet{Kri_etal03} to obtain $BVRI$ S-corrections;
these spectra extend to 2950 \AA\ and are shown and discussed by \citet{Fol_etal08b}.
For SN 2002bo we used the four spectra of \citet{Ben_etal04} that extended
to 3200 \AA\ in the blue or beyond.  For SN 2004S we used two of the 
spectra discussed by \citet{Kri_etal07}. 

As found by \citet{Kri_etal11} and \citet{Bar_etal12}, SN 2001ay is an unusual
object, and perhaps the prototype of a new subclass of Type Ia SNe.  It
is the most slowly declining Type Ia SN found to date, but it has
a normal peak luminosity.  Its photometric behavior can be understood in terms of a 
pulsating delayed detonation model, which gives a relatively rapid
rise to maximum light and a very slow decline.  These objects are
undoubtedly very rare.  In Figure \ref{u_scorr_01ay} we show the
$U$-band S-corrections of SN 2001ay, which are numerically quite different
than the corrections based on spectra of four ``normal'' objects shown 
in Figure \ref{u_scorr}.

\section {Newly published data}

In Table \ref{u_2003gs} we give fully corrected $U$-band photometry of
SN 2003gs obtained with the Las Campanas Observatory 1-m telescope.
Previously, \citet{Kri_etal09} published $U$-band photometry without 
S-corrections from observations with the CTIO 1.3- and 0.9-m 
telescopes.\footnote[8]{Extensive optical spectra of SN 2003gs 
(Kotak et al., in preparation) were not available to us for the calculation of
S-corrections.}  Figure \ref{03gs_corr} shows the fully corrected $U$-band 
light curve of SN 2003gs from observations made with these three telescopes.
This is a significant improvement compared to the light curve shown in
the upper panel of Figure \ref{u_uncorrected}.  The
rms scatter of the $U$-band photometry is $\pm$ 0.085 mag.

In Table \ref{ubvri} we give S-corrected photometry of SN 2003hv based on 
15 nights of observations with the CTIO 1.3-m telescope and ANDICAM.  In addition
to the field stars used by \citet{Lel_etal09} we needed an extra
secondary standard; it is located at right ascension $\alpha$ =
3:04:00.3, declination $\delta$ = $-$26:08:43 (J2000).  From observations
on four photometric nights in September and October of 2003, we find
$U$= 14.187, $B$ = 14.124, $V$ = 13.577, $R$ = 13.234, $I$ = 12.926
for this star, with internal random errors of $\pm$ 0.005 mag.
Without this extra star, which is reasonably bright in the $U$-band,
we could not have accurately calibrated the CTIO 1.3-m $U$-band data
obtained on non-photometric nights.

Figure \ref{03hv_corr} shows the S-corrected light curves of SN 2003hv
based on data from the LCO 1-m, CTIO 1.3- and 0.9-m telescopes.  
\citet{Lel_etal09} previously published data without the S-corrections.
G. Leloudas kindly provided the $BVRI$ data with the S-corrections.
Now the $U$-band light curve of SN 2003hv is greatly improved 
compared to the light curve shown in the lower panel of
Figure \ref{u_uncorrected}.  The rms scatter of the $U$-band photometry
is $\pm$ 0.074 mag.

SNe 2003gs and 2003hv were reasonably fast decliners, with 
\dmm\ = 1.83 and 1.61 mag, respectively. As 
a result, we would expect them to be subluminous in optical passbands and 
intrinsically red. Figure \ref{03hv_ub} shows the $U-B$ colors of SNe 2003gs and 
2003hv.  For  SN 2003gs, E($U-B$) = 0.025 mag.  This fast declining object 
clearly had much  redder $U-B$ colors than SN~2003hv.
SN~2003hv was unreddened in its host, and we have subtracted off 
a small amount of reddening due to Milky Way dust (E($U-B$) = 0.011 mag), based on 
the reddening maps of \citet{Sch_etal98}, as corrected by \citet{Sch_Fin11}.  

We may compare the $U-B$ color curves of SNe 2003gs and 2003hv to
loci shown in  Figure 45 of 
\citet{Kes_etal09}.  They show color curves for three values of the 
Multi-color Light Curve Shape parameter $\Delta$.\footnote[9]{The 
MLCS parameter $\Delta$ is effectively the number of $V$-band magnitudes that
a Type I SN at maximum light is brighter or fainter than an object of
nominal decline rate.  Slowly rising, slowly declining
objects, which are overluminous, have $\Delta < 0$.  $\Delta$ ranges from
about $-$0.4 to +1.4.} 
Figure 45 of \citet{Kes_etal09} also shows loci modified to reproduce
the flux and color evolution of objects from the SNLS project.
To help decide which light curve fitting model is best for fitting
restframe $U$-band data of high redshift objects we need the corrected photometry of
a sample of nearby slow decliners.  Magnitude limited surveys discover a
high percentage of such Type Ia SNe.

In Figure \ref{03hv_ub} we show three loci from the nominal Multicolor
Light Curve Shape model \citep[MLCS2K2,][]{Jha_etal07}.  Of the three loci
shown in this figure the dereddened colors of SN~2003hv are best fit
by the $\Delta$ = +0.4 locus prior to $t$ = 20 d.  

\section{Discussion}

The future of ground based photometry of supernovae will involve end-to-end
calibration of whole telescope-plus-camera systems.  A review of the concerns
involved is given by \citet{Stu_Ton06}.  \citet{Stu_etal07} describe
preliminary results with a tunable laser system used on the CTIO 4-m
telescope.  \citet{Rhe_etal10} describe a {\em monochromator} that measures
the throughput of a filter in a camera on a telescope at a range of
wavelengths simultaneously; this system works from short wavelength
optical light into the near-IR.  The purpose of these systems is to
eliminate the guesswork involved in multiplying together the
wavelength dependent atmospheric transmission, the
optical reflection and transmission functions, and the quantum
efficiency of the detector, in an attempt
to determine the effective transmission profiles of a telescopic system.

Our $U$-band results presented here were not produced with a
system that measures the filter transmission {\em in situ}.
But it is clear from Figures \ref{03gs_corr} and \ref{03hv_corr} that the
application of S-corrections done the traditional way gives greatly
improved $U$-band light curves in the cases of SNe 2003gs and 2003hv.
We found values of the rms scatter of the $U$-band photometry equal
to $\pm$ 0.085 and 0.074 mag, respectively, for these two objects.
This is comparable to the scatter ($\pm$ 0.077 mag) 
of S-corrected Sloan $u$-band photometry of Type Ia SNe 
observed with multiple telescopes \citep{Mos_etal12}.

We can apply our S-corrections to the $U$-band
light curves of two other normal objects, SNe 2001el and 2004S.
SN 2001el was observed in the $U$-band at maximum light using
the CTIO 0.9-m telescope and thereafter with the CTIO 1.5-m telescope
using an essentially identical $U$-band filter.
SN 2004S was observed in the $U$-band with the CTIO 1.3-m telescope
and ANDICAM.

\citet{Kri_etal07} found that SNe 2001el and 2004S were essentially 
clones of each other. These two objects suffered known small 
amounts of interstellar extinction from 
dust in the Milky Way, but SN 2001el suffered more host galaxy
extinction.  Figure 14 of \citet{Kri_etal07} shows the {\em difference}
of the broad band apparent magnitudes of the two objects, corrected
for the effect of Milky Way dust.  But the $U$-band point was anomalous.  
The corrected $U$-band point required a shift of 0.35 mag.  Now 
all the points can be fit very well by a simple curve (see Figure \ref{dmag2}).
Such data allow us to show that a combination of optical and
near-IR data gives us the most accurate values of R$_V$ and 
A$_V$.\footnote[10]{Here A$_V$ = R$_V$ E($B-V$), where A$_V$ is
the  $V$-band interstellar extinction, E($B-V$) is the color excess 
for $B-V$ colors, and R$_V$ is a scale factor.  See \citet{Car_etal89}
and \citet{Kri_etal06}.}  

Another possibility is that while the \citet{Car_etal89} extinction law 
(CCM89) might work well to account for dust extinction in the Milky Way, it 
might actually not be the right extinction model to account for the 
extinction due to circumstellar dust and interstellar dust that affects the 
light of Type Ia SNe in other galaxies (Shappee \& Jha, in preparation), at 
least in the UV. Our corrected $U$-band point in Fig. \ref{dmag2} is 
1.8-$\sigma$ different than what we would expect using a CCM89 extinction 
law with $R_V$ = 2.15, the most robust value derived from $BVRIJHK$ 
photometry.

We note that the mean $U$-band S-correction for the CTIO 1.3-m telescope
to convert the data to the \citet{Bes90} filter prescription
is $\Delta U \approx -0.21$ mag for four ``normal'' objects 
(after the elimination of two outlying points).  For the CTIO 0.9-m the 
mean value is about +0.16 mag.  For the LCO 1-m telescope the mean value is 
about +0.25 mag.  While the determination of S-corrections is non-trivial,
we note that the determination of the mean color term in Equation 1 is
easily done with observations of a dozen or more standard stars
covering a range of color, obtained on a number of photometric nights. 

In Figure \ref{ct_corr} we show a sparse empirical plot of the $U$-band S-correction
at 3 days after $B$-band maximum vs. the $U$-band color term for the
three telescope systems we are considering here.  The bluest $U$-band filter
(for the CTIO 1.3-m telescope) leads to the largest negative color term
and the largest negative S-correction.  The reddest $U$-band filter (for
the LCO 1-m telescope) leads to the largest positive color term and
the largest positive S-correction.  It would certainly be worthwhile to
use the effective $U$-band filter transmission curves from other systems
to derive S-corrections to see to what extent the linear trend shown
in Figure \ref{ct_corr} holds.  

We surmise that the resolution of the $U$-band anomaly will require
multiple upgrades to the analysis: 1) improved effective filter
profiles, allowing better standardization of the data; 2) adoption
of a reddening law that differs from a CCM89-type extinction law
tied to a specific value of R$_V$, at least in the UV; and 3) 
retraining the lightcurve fitters to account for evolution in
the UV spectral energy distributions of Type Ia SNe as a function
of redshift \citep {Fol_etal12, Mag_etal12}.
When it comes to determining the equation 
of state parameter of the universe, $w = P/(\rho c^2)$, we do not want 
attribute to the universe something that is in a subtle way an artifact 
of data analysis pushed to its limits.

\section{Conclusions}

Over the past dozen years we have published $U$-band photometry of
quite a few Type Ia SNe, but we have never really done anything
with the $U$-band data.  It was obvious that there were systematic offsets 
from telescope to telescope. We previously opted not to publish $U$-band
photometry of SN 2003gs from the LCO 1-m telescope \citep{Kri_etal09}.
Here we have worked out S-corrections for that
telescope, along with corrections for the CTIO 0.9-m and 1.3-m telescopes.
Even though we did not measure the effective $U$-band filter profiles
using end-to-end calibration, our S-corrections provide a significant
improvement to the analysis.

\citet{Kes_etal09} discovered that if we anchor the absolute magnitudes
of distant Type Ia SNe (redshift $z \gtrsim$ 0.3) with $U$-band light curves
of nearby objects, there is a 0.12 mag discrepancy in the distance moduli,
which leads to a 0.3 shift in the equation of state parameter $w$.
Using WMAP data, baryon acoustic oscillations and
Type Ia supernovae, the random error of $w$ from modern surveys is now 
about $\pm$ 0.06 \citep{Kom_etal11}, and within one standard deviation 
it appears that $w$ = $-$1.\footnote[11]{If 
$w$ = $-$1, then the Dark Energy is equivalent to Einstein's Cosmological
Constant.}  But a systematic error of 0.3 is so serious that it ruins the 
experiment.

Our experiments with traditional S-corrections presented in this 
paper should give us hope and motivate us to compile a set of S-corrected
light curves of nearby objects that cover maximum light as well as possible.
This should better enable the full use of restframe $U$-band photometry of
higher redshift objects.  We could use these corrected light curves to
retrain the light curve fitters used for such projects as ESSENCE, 
SDSS, and SNLS and see if the $U$-band anomaly found by 
\citet{Kes_etal09} is resolved.  This may be time well spent while
new projects such as the Dark Energy Survey begin operations.

\acknowledgments

We thank Kyle Owens and Jude Magaro for help reducing some of 
the photometry of SN 2003hv.  Giorgos Leloudas kindly provided
S-corrected $BVRI$ photometry of SN 2003hv, the uncorrected photometry
of which was presented in \citet{Lel_etal09}.  We thank Max Stritzinger,
Alex Conley, Mark Sullivan, Andy Becker, Rick Kessler, Peter Brown,
Saurabh Jha, Malcolm Hicken, and an anonymous referee for 
useful comments.


\clearpage

\begin{deluxetable}{lcc}
\tablewidth{0pt}
\tablecolumns{8}
\tablecaption{Fully Corrected $U$-Band Photometry of SN~2003gs\label{u_2003gs}}
\tablehead{
\colhead{JD\tablenotemark{a}} &
\colhead{Epoch\tablenotemark{b}} &
\colhead{$U$} 
}
\startdata
2850.80   & \phn+1.99    &  14.566 (0.018)  \\ 
2851.80   & \phn+2.99    &  14.647 (0.018)  \\        
2870.80   & +21.89       &  17.026 (0.018)  \\      
2888.90   & +39.91       &  17.575 (0.018)  \\    
2890.80   & +41.80       &  17.603 (0.018)  \\
2905.70   & +56.63       &  18.192 (0.020)  \\  
2906.80   & +57.72       &  18.177 (0.023)  \\
2907.80   & +58.72       &  18.197 (0.019)  \\
2908.80   & +59.71       &  18.265 (0.021)  \\
2914.80   & +65.68       &  18.405 (0.040)  \\
\enddata
\tablenotetext{a} {Julian Date {\em minus} 2,450,000.}
\tablenotetext{b} {Restframe days since T($B_{max}$) = JD 2452848.8.}
\end{deluxetable}

\begin{deluxetable}{lcccccc}
\tablewidth{0pt}
\tabletypesize{\scriptsize}
\tablecolumns{7}
\tablecaption{Fully Corrected Optical Photometry of SN~2003hv\tablenotemark{a}\label{ubvri}}
\tablehead{
\colhead{JD\tablenotemark{b}} &
\colhead {Epoch\tablenotemark{c}} &
\colhead{$U$} &
\colhead{$B$} &
\colhead{$V$} &
\colhead{$R$} &
\colhead{$I$} 
}
\startdata
2892.88 &  \phn+1.17  & 11.976 (0.015) & 12.416 (0.015) & 12.480 (0.015) & 12.425 (0.015) & 12.756 (0.015) \\
2896.78 &  \phn+5.05  & 12.512 (0.015) & 12.674 (0.015) & 12.607 (0.015) & 12.612 (0.015) & 12.998 (0.015) \\
2899.79 &  \phn+8.04  & 12.967 (0.017) & 13.011 (0.015) & 12.798 (0.015) & 12.908 (0.015) & 13.273 (0.015) \\
2906.80 &  +15.02     & 14.184 (0.015) & 14.012 (0.015) & 13.299 (0.015) & 13.181 (0.015) & 13.116 (0.015) \\
2910.76 &  +18.95     & 14.677 (0.015) & 14.577 (0.041) & 13.573 (0.034) & 13.317 (0.035) & 13.016 (0.037) \\
2914.70 &  +22.87     & 15.054 (0.032) & 15.024 (0.066) & 13.954 (0.033) & 13.621 (0.034) & 13.191 (0.052) \\
2917.75 &  +25.90     & 15.233 (0.015) & 15.180 (0.015) & 14.141 (0.015) & 13.825 (0.015) & 13.427 (0.015) \\
2920.74 &  +28.88     & 15.408 (0.025) & 15.290 (0.029) & 14.363 (0.024) & 14.065 (0.031) & 13.708 (0.037) \\
2923.72 &  +31.84     & 15.501 (0.022) & 15.403 (0.015) & 14.493 (0.015) & 14.220 (0.015) & 13.895 (0.015) \\
2926.75 &  +34.85     & 15.609 (0.022) & 15.507 (0.015) & 14.626 (0.015) & 14.367 (0.015) & 14.078 (0.015) \\
2929.72 &  +37.81     & 15.755 (0.015) & 15.579 (0.015) & 14.729 (0.015) & 14.453 (0.015) & 14.210 (0.015) \\
2932.67 &  +40.74     & 15.810 (0.015) & 15.656 (0.015) & 14.821 (0.015) & 14.595 (0.015) & 14.385 (0.015) \\
2939.71 &  +47.74     & 16.042 (0.024) & 15.827 (0.015) & 15.044 (0.015) & 14.851 (0.015) & 14.726 (0.025) \\
2946.71 &  +54.70     & 16.197 (0.029) & 15.957 (0.015) & 15.236 (0.015) & 15.102 (0.015) & 15.033 (0.015) \\
2953.72 &  +61.67     & 16.301 (0.058) & 16.105 (0.018) & 15.435 (0.020) & 15.343 (0.015) & 15.336 (0.015) \\
\enddata
\tablenotetext{a}{The photometry includes the usual color corrections
derived from standard stars, and also the S-corrections. 
Magnitude uncertainties ($1\sigma$) are given in parentheses.}
\tablenotetext{b}{Julian Date {\em minus} 2,450,000.}
\tablenotetext{c}{Rest frame days since T($B_{max}$ = JD 2452891.7).}
\end{deluxetable}

\clearpage

\figcaption[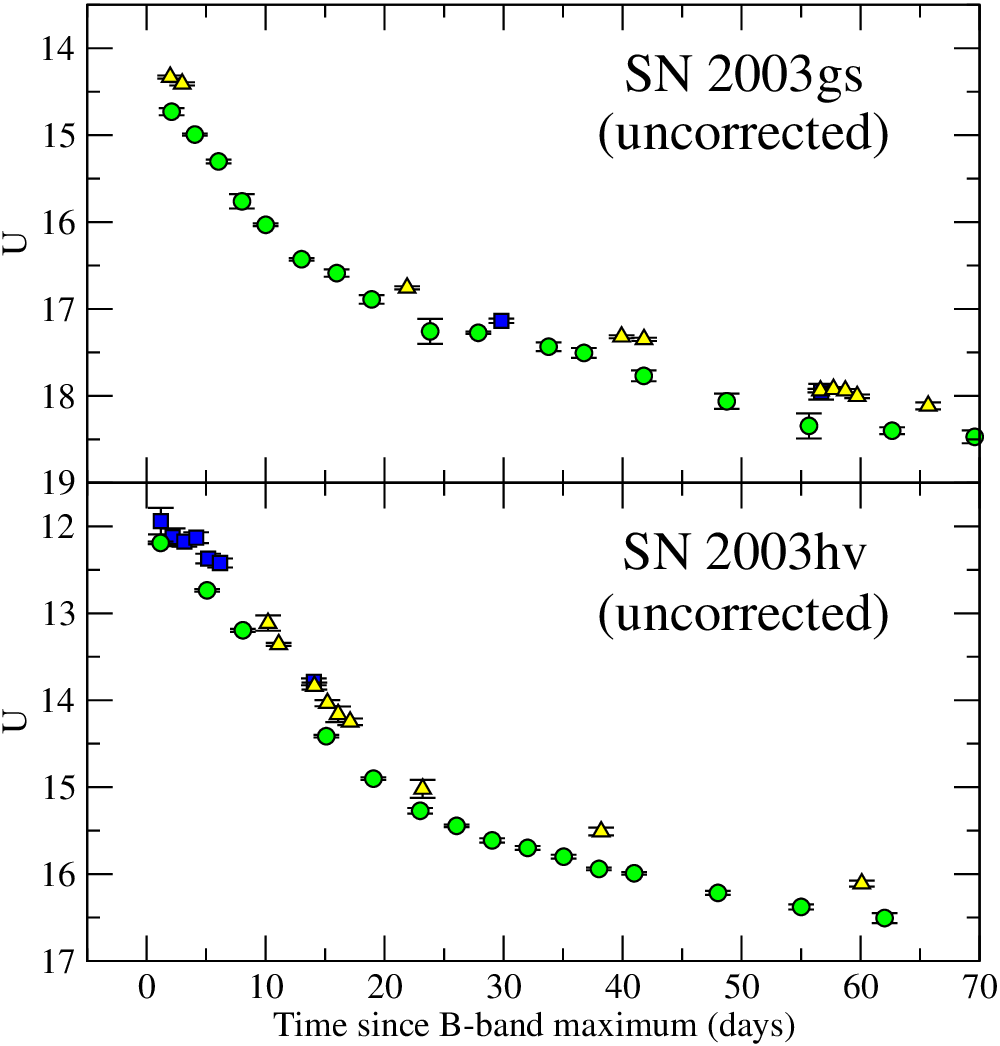] {$U$-band photometry (without
S-corrections) of SN~2003gs and SN~2003hv.  Symbols: green circles =
CTIO 1.3-m with ANDICAM; blue squares = CTIO 0.9-m; yellow triangles =
LCO 1-m.
\label{u_uncorrected}
}

\figcaption[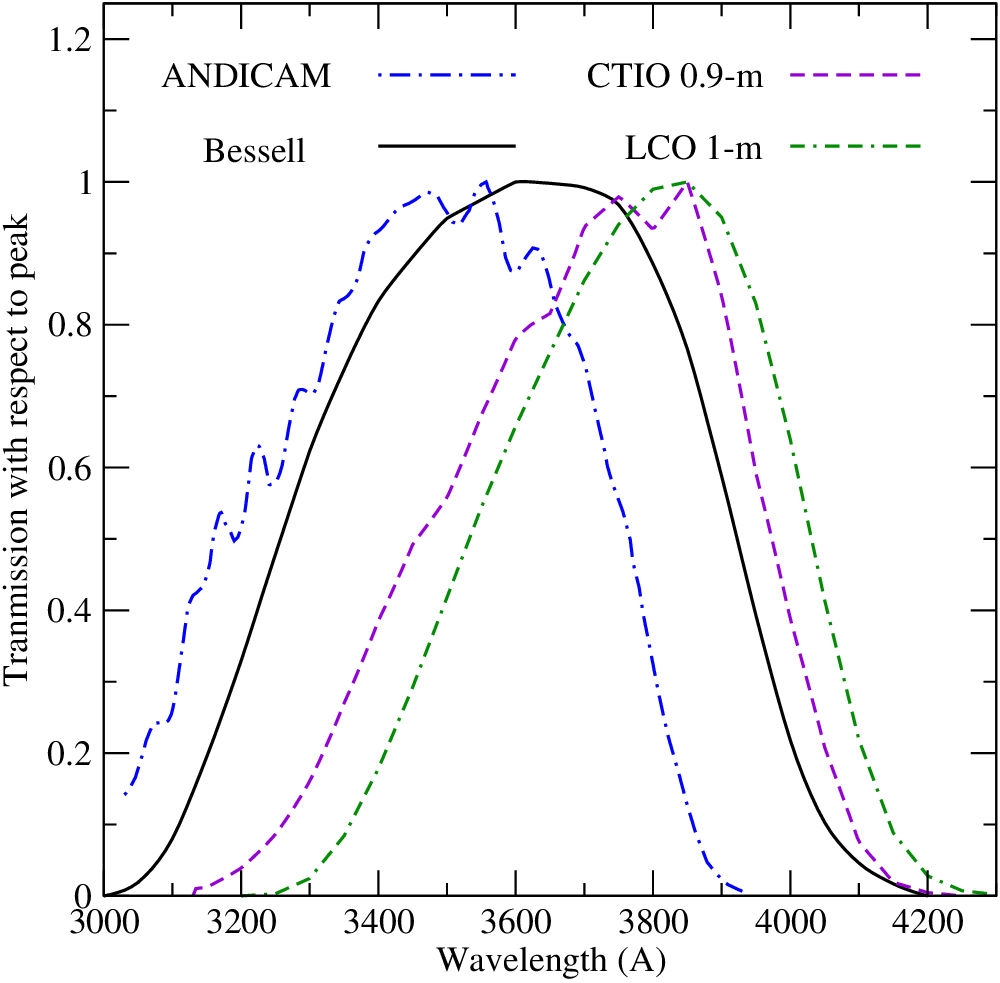] {Effective $U$-band transmission functions
of the cameras on the CTIO 1.3-m, CTIO 0.9-m, and LCO 1-m telescopes,
along with the $U$-band filter specified by \citet{Bes90}. 
\label{u_trans}
}

\figcaption[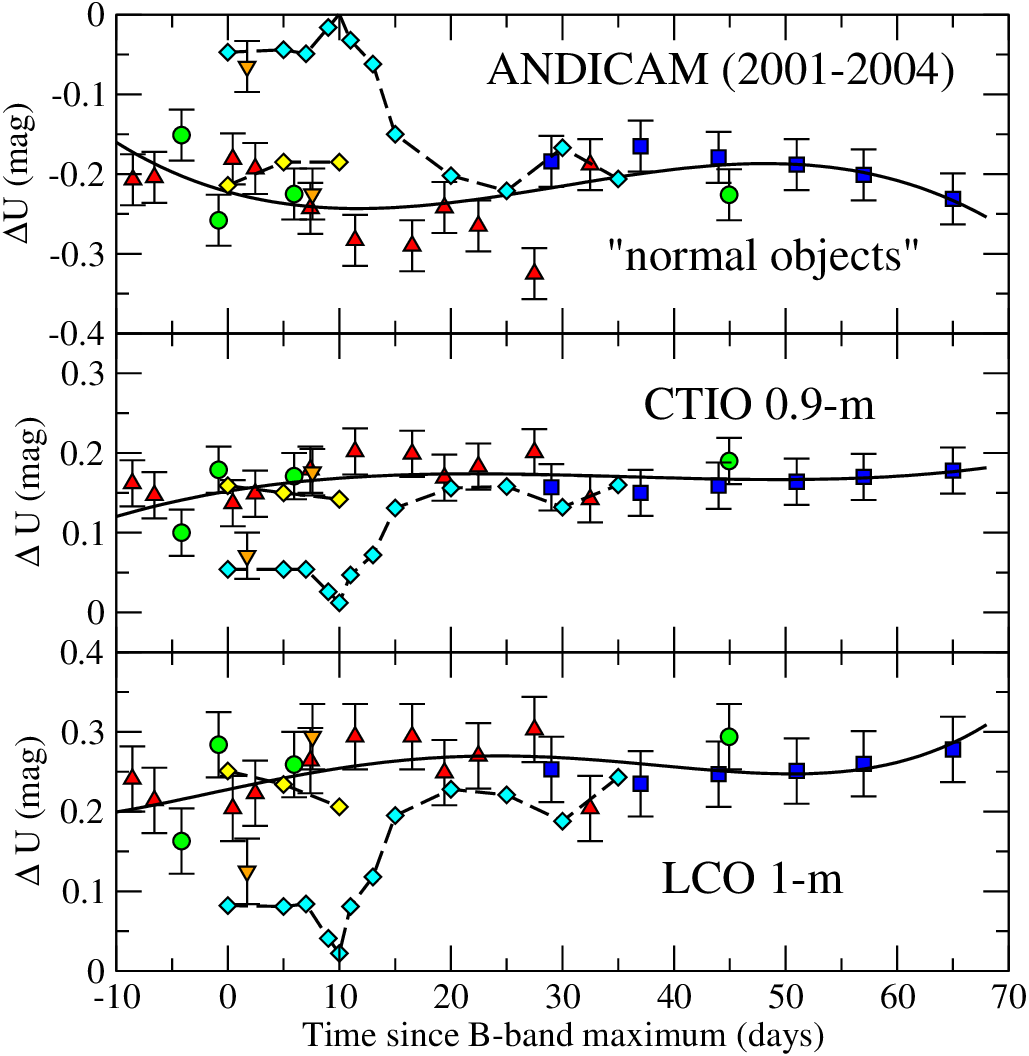] {$U$-band S-corrections for 
``spectroscopically normal objects''.  Symbols: red upward pointing triangles =
SN~1999ee; blue squares = SN~2001el; green dots = SN~2002bo; orange
downward pointing triangles = SN~2004S.  The top panel shows corrections for
the ANDICAM instrument on the CTIO 1.3-m telescope.  The middle panel shows
corrections for the CTIO 0.9-m reflector.  The bottom panel shows corrections
for the Las Campanas Observatory 1-m telescope. The solid lines are low order
polynomial fits to these data.  Connected by dashed lines we also show 
S-corrections based on \citet {Hsi_etal07} templates warped to match the
natural system magnitudes of SN~1999ee data from the CTIO 0.9-m telescope
(cyan diamonds) and SN~2002bo data from ANDICAM when it was mounted on
the Yale-AURA-Lisbon-Ohio (YALO) telescope at CTIO (yellow diamonds).
\label{u_scorr}
}

\figcaption[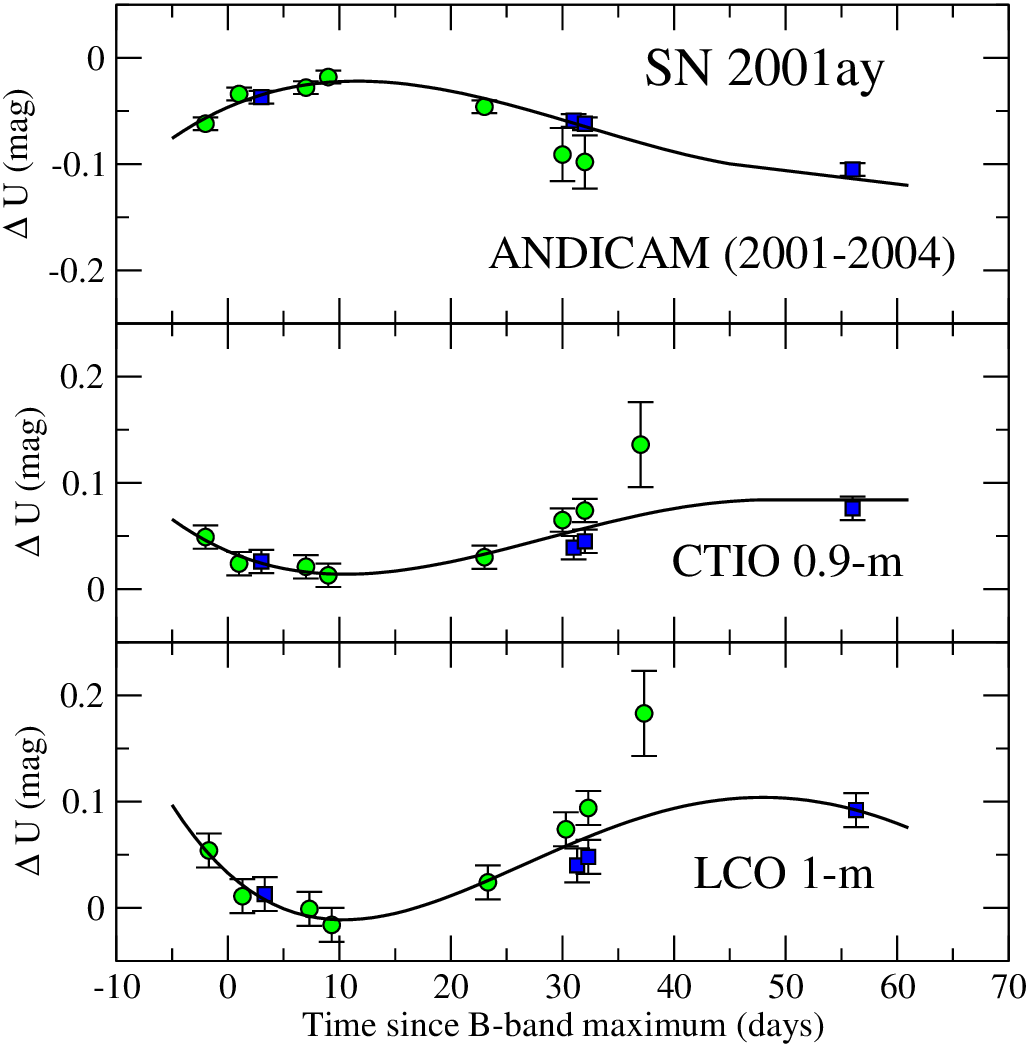] {Similar to Figure \ref{u_scorr},
except these are $U$-band S-corrections for the unusal Type Ia
SN~2001ay.  The green dots correspond to corrections based on spectra
with the Fred L. Whipple Observatory 1.5-m telescope.  The blue squares
correspond to corrections base on spectra with the MMT.
\label{u_scorr_01ay}
}

\figcaption[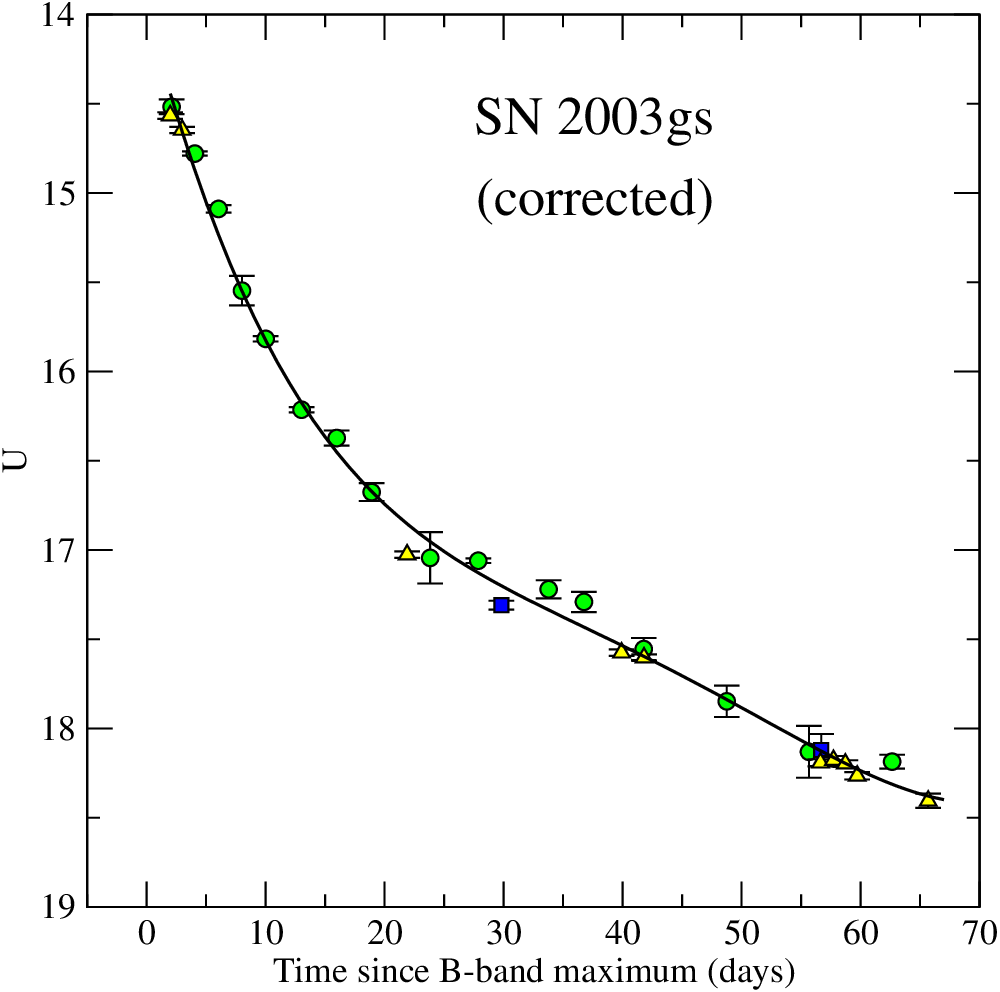] {$U$-band photometry (with
S-corrections) of SN~2003gs.  There are now no significant
systematic differences of photometry obtained with different
telescopes.
\label{03gs_corr}
}

\figcaption[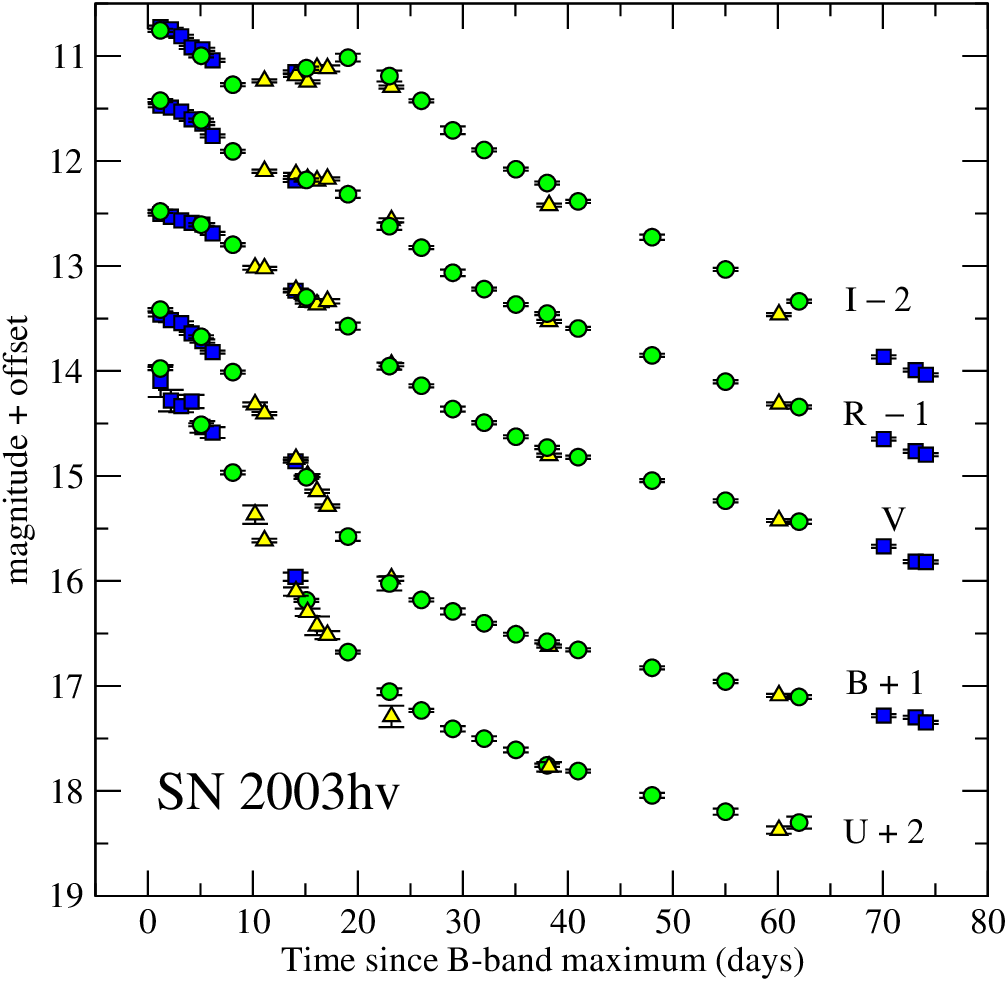] {$UBVRI$-band photometry (with
S-corrections) of SN~2003hv.  The CTIO 0.9-m
and LCO 1-m $BVRI$ data (blue squares and yellow upward pointing 
triangles, respectively) are from \citet{Lel_etal09}, with S-corrections
provided by G. Leloudas.  The CTIO 1.3-m data are plotted
as green dots.  We have excluded KAIT data and 
data from the Siding Spring Observatory 2.3-m.  
\label{03hv_corr}
}

\figcaption[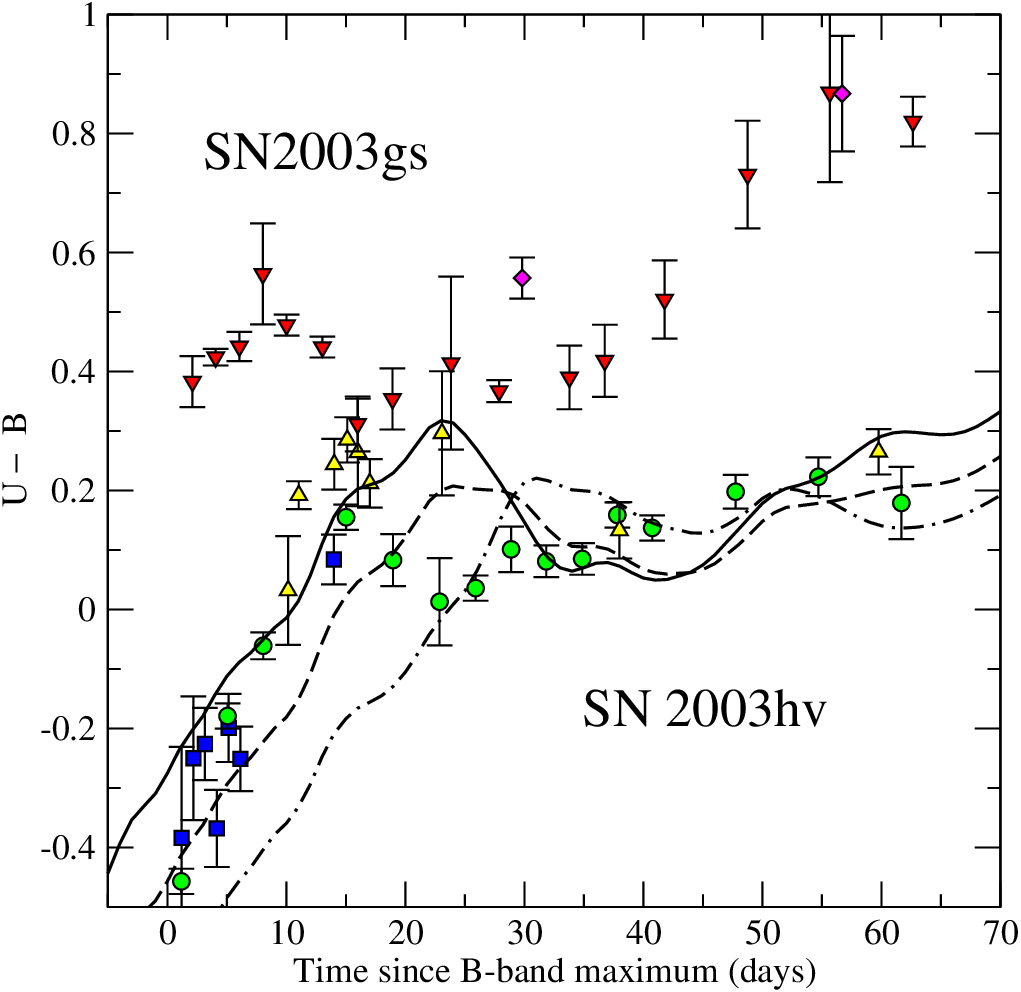] {$U-B$ color curves of 
SNe 2003hv and SN 2003gs.  The photometry of SN 2003hv
has been S-corrected and
corrected for a small amount of Milky Way reddening
(E($U-B$) = 0.011).  Symbols: green dots (CTIO 1.3-m),
blue squares (CTIO 0.9-m), yellow upward pointing triangles (LCO 1-m).
For SN 2003gs, E($U-B$) = 0.025 mag, which has been subtracted out.
Symbols: red downward pointing triangles (CTIO 1.3-m),
magenta diamonds (CTIO 0.9-m).  The solid line is the MLCS2K2 locus
for $\Delta$ = +0.4.  The dashed line is for $\Delta$ = 0.0.  The
dot-dash line is for $\Delta = -0.4$.
\label{03hv_ub}
}

\figcaption[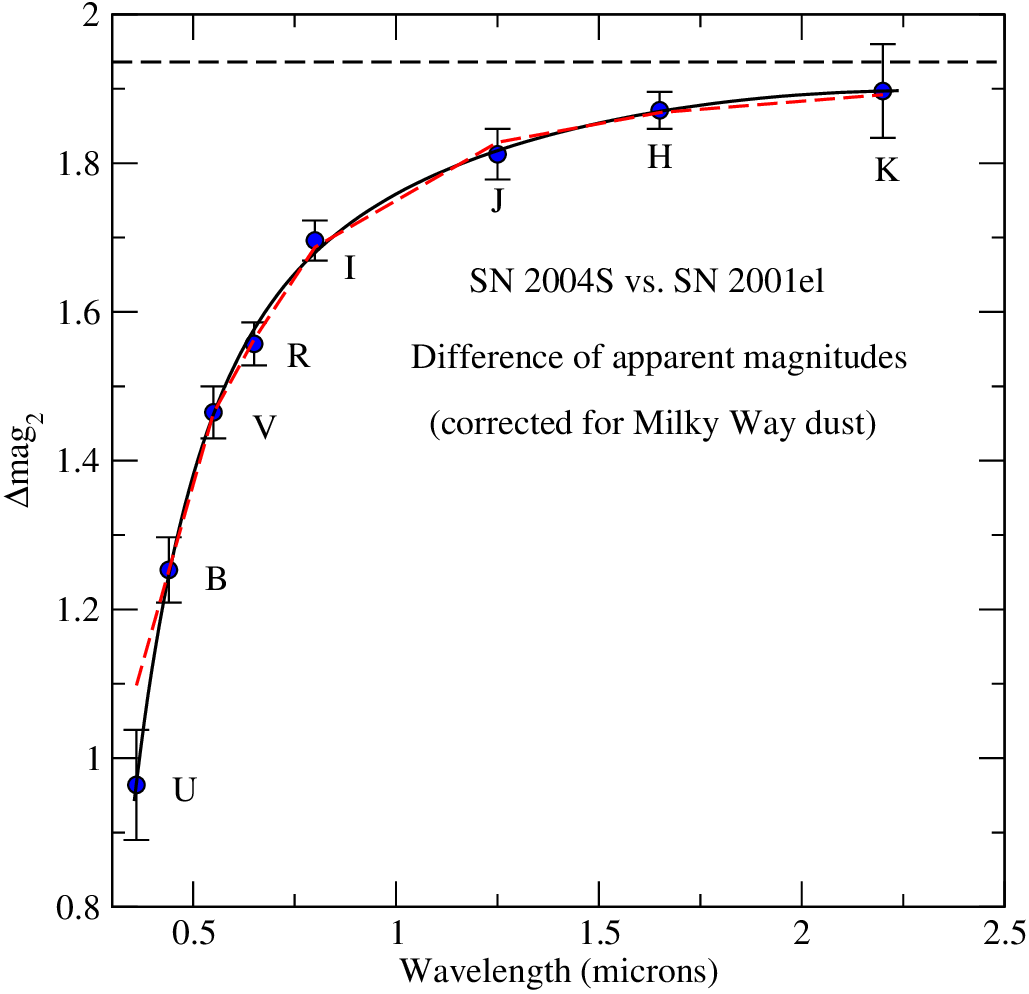] {Difference of apparent magnitudes
of SNe 2001el and 2004S, following Figure 14 of \citet{Kri_etal07}.
The $U$-band difference is now based on S-corrected photometry.
The horizontal dashed line is the derived difference of the distance moduli.
The solid line is a fourth order fit to the $UBVRIJHK$ points.  The
red dashed line is based on a distance modulus difference of 1.936 mag,
$A_V$ = 0.472 mag and scale factors $A_{\lambda}/A_V$ from Table 6 
of \citet{Kri_etal07}, using a CCM89 extinction law and $R_V$ = 2.15.
\label{dmag2}
}

\figcaption[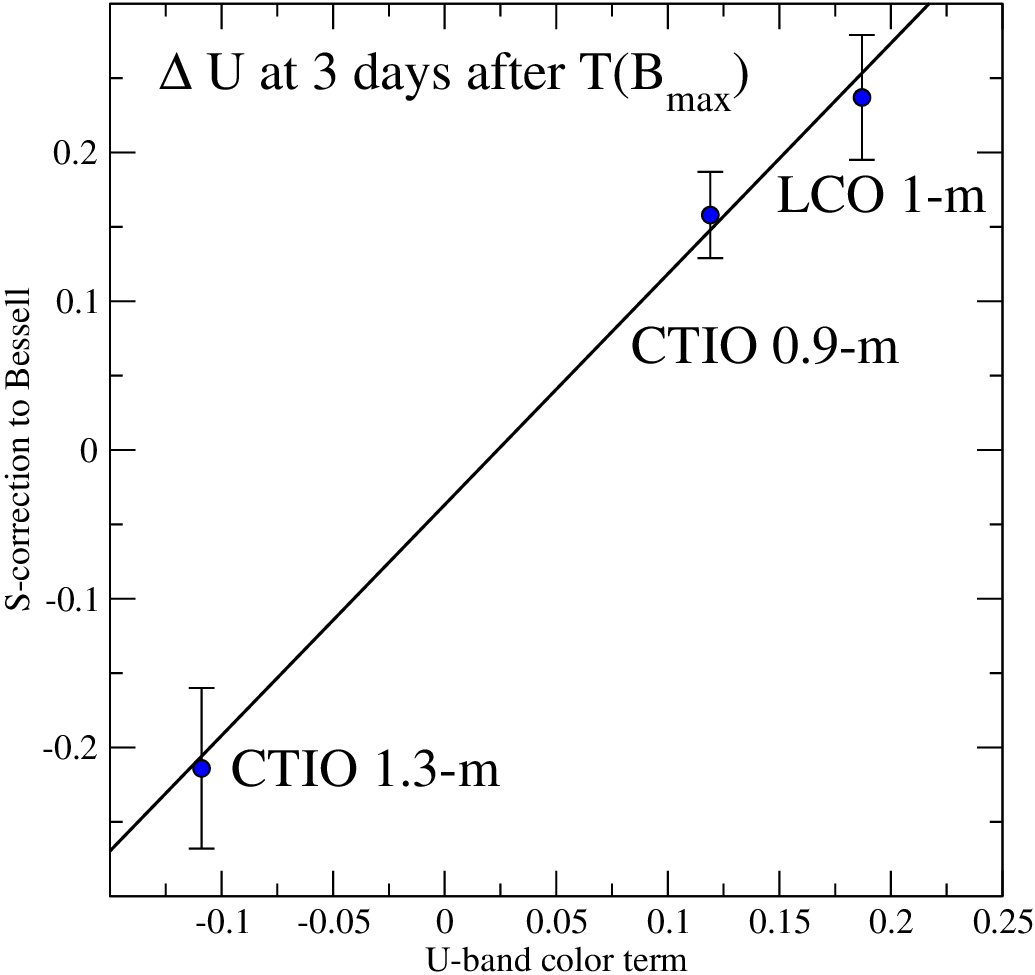] {$U$-band S-correction as a function
of $U$-band photometric color term.
\label{ct_corr}
}

\clearpage

\begin{figure}
\plotone{u_uncorrected.eps}
{\center Krisciunas {\it et al.} Fig. \ref{u_uncorrected}}
\end{figure}

\begin{figure}
\plotone{u_trans.eps}
{\center Krisciunas {\it et al.} Fig. \ref{u_trans}}
\end{figure}

\begin{figure}
\plotone{u_scorr.eps}
{\center Krisciunas {\it et al.} Fig. \ref{u_scorr}}
\end{figure}

\begin{figure}
\plotone{u_scorr_01ay.eps}
{\center Krisciunas {\it et al.} Fig. \ref{u_scorr_01ay}}
\end{figure}

\begin{figure}
\plotone{03gs_corrected.eps}
{\center Krisciunas {\it et al.} Fig. \ref{03gs_corr}}
\end{figure}

\begin{figure}
\plotone{sn2003hv_ubvri.eps}
{\center Krisciunas {\it et al.} Fig. \ref{03hv_corr}}
\end{figure}

\begin{figure}
\plotone{sn2003hv_gs.eps}
{\center Krisciunas {\it et al.} Fig. \ref{03hv_ub}}
\end{figure}

\begin{figure}
\plotone{dmag2.eps}
{\center Krisciunas {\it et al.} Fig. \ref{dmag2}}
\end{figure}

\begin{figure}
\plotone{ct_corr.eps}
{\center Krisciunas {\it et al.} Fig. \ref{ct_corr}}
\end{figure}

\end{document}